\documentclass[12pt,epsf,qsymbols]{article}
\pdfoutput=1
\usepackage[utf8]{inputenc}
\usepackage[normalem]{ulem}
\usepackage[english]{babel}%,frenchb
\usepackage{tabularx}
\usepackage{array}
\usepackage{graphics}
\usepackage[pdftex]{graphicx}
\usepackage{psfrag}
\usepackage{epsfig}
\usepackage{subfig}
\usepackage{mathtools}
\usepackage{amssymb}
\usepackage{bbold}
\usepackage{setspace}
\usepackage{rotating}
\usepackage{colortbl}
\usepackage{longtable}
\usepackage{slashed}
\usepackage{braket}
\usepackage{lineno}
\makeatletter
\usepackage{textcomp}
\usepackage[usenames,dvipsnames]{xcolor}
\usepackage{relsize}
\usepackage{cite}

\usepackage{float}

\usepackage{bbm}
\usepackage{bm}
\usepackage{enumerate}
\usepackage{amsmath}
\usepackage{verbatim}

\usepackage{hyperref}
\hypersetup{colorlinks,citecolor=nicegreen,linkcolor=niceblue}
\hypersetup{colorlinks=true}

\usepackage{pifont}% http://ctan.org/pkg/pifont

\allowdisplaybreaks

\setlongtables

\newcommand{\bea}{\begin{eqnarray}}
\newcommand{\eea}{\end{eqnarray}}
\newcommand{\beq}{\begin{equation}}
{
\newcommand{\eeq}{\end{equation}}
\newcommand{\ec}{\end{center}}
\newcommand{\bc}{\begin{center}}

\newcommand{\pdir}{p\kern -5.2pt\raise 0.2ex\hbox {/}}

\newcommand{\vdir}{v\kern -5.75pt\raise 0.15ex\hbox {/}}
\newcommand{\kdir}{k\kern -5.75pt\raise 0.15ex\hbox {/}}
\newcommand{\epsdir}{\epsilon\kern -5.0pt\raise 0.15ex\hbox {/}}
\newcommand{\bvdir}{\bar{v}\kern -5.75pt\raise 0.15ex\hbox {/}}
\newcommand{\Ddir}{D\kern -7.75pt\raise 0.20ex\hbox {/}}
\newcommand{\Adir}{A\kern -7.75pt\raise 0.20ex\hbox {/}}
\newcommand{\ldir}{l\kern -5.0pt\raise 0.2ex\hbox{/}}
\newcommand{\varepsdir}{\varepsilon\kern -5.5pt\raise 0.15ex\hbox{/}}

\makeatother

\definecolor{niceblue}{rgb}{0.15,0.15,0.6}
\definecolor{nicegreen}{rgb}{0.1,0.5,0.1}
\definecolor{Red}{rgb}{1.,0.,0.}

\definecolor{Green}{rgb}{0.2,.7,0.2}

\begin{document}
\unitlength = 1mm

\thispagestyle{empty} 
\begin{flushright}
\begin{tabular}{l}
%{\tt \footnotesize UNL-XX-XX}\\
\end{tabular}
\end{flushright}
\begin{center}
\vskip 3.4cm\par
{\par\centering \textbf{\LARGE  
\Large \bf Multistep Strongly First Order Phase Transitions  \\[0.3em] from New Fermions at the TeV Scale}}
\vskip 1.2cm\par
{\scalebox{.85}{\par\centering \large  
\sc Andrei Angelescu$^a$, Peisi Huang$^a$}
{\par\centering \vskip 0.7 cm\par}
{\sl 
$^a$~{Department of Physics and Astronomy\\ University of Nebraska-Lincoln, Lincoln, NE, 68588, USA.}}\\

{\vskip 1.65cm\par}}
\end{center}

\vskip 0.85cm
\begin{abstract}

In spite of the vast literature on the subject of first order Electroweak Phase Transitions (EWPT), which can provide the necessary conditions for generating the Baryon Asymmetry in the Universe, fermion-induced EWPTs still remain a rather uncharted territory. In this paper, we consider a simple fermionic extension of the Standard Model involving one $SU(2)_L$ doublet and two $SU(2)_L$ singlet Vector-Like Leptons, strongly coupled to the Higgs scalar and with masses close to the TeV scale. We show how such a simple scenario can give rise to a non-trivial thermal history of the Universe, involving strongly first order multistep phase transitions occurring at temperatures close to the electroweak scale. Finally, we investigate the distinct Gravitational Wave (GW) signatures of these phase transitions at future space--based GW detectors, such as LISA, DECIGO, and BBO, and briefly discuss the possible LHC signatures of the VLLs.

\end{abstract}
\newpage
\setcounter{page}{1}
\setcounter{footnote}{0}
\setcounter{equation}{0}
%%%%%%%%%%%%%%%%%%%%%%%%%%%%%%%%%%%%%%%%
\noindent

\renewcommand{\thefootnote}{\arabic{footnote}}
%\linenumbers

\setcounter{footnote}{0}

\tableofcontents

\newpage

%%%%%%%%%%%
%%%%%%%%%%%
%%%%%%%%%%%
%\section{Introduction}
%\label{sec:intro}

\section{Introduction} \label{sec:intro}

The origin of the baryon asymmetry in the Universe (BAU), which is also known as the matter-antimatter asymmetry, is one of the most outstanding problems in modern cosmology. %One can assume that the asymmetry was an initial condition of the Universe, but such a solution seems ad--hoc and would suffer from a naturalness problem. Instead, a dynamical origin of the BAU, dubbed baryogenesis, would be a more appealing scenario. 
Baryogenesis is an appealing scenario of generating the matter-antimatter asymmetry dynamically. In 1967, Sakharov enunciated three conditions for a successful baryogenesis: baryon number violation, $C$ and $CP$ violation, and a departure from the thermal equilibrium~\cite{sakharov:1967dj}.

Of particular interest for this work is the third condition, departure from thermal equilibrium, which can only be met if the Universe underwent a strongly first order phase transition (SFOPT) in its early stages. While there is no indication about the energy scale at which it happened, an attractive possibility would be that such a phase transition (PT) occurred around the Electroweak (EW) scale. Indeed, it is beyond any doubt that EW symmetry was broken at some point in the history of the Universe, and a straightforward possibility would be that the PT responsible for baryogenesis took place when EW symmetry was broken. 

As any cosmological PT, an Electroweak phase transition (EWPT) would generate a stochastic GW background sourced in the collision of nucleating bubbles and the plasma motion induced by bubble collisions. In the specific case of a strong EWPT, it is expected that, due to redshift, the GW signal would nowadays peak at frequencies around the mHz-dHz range\textcolor{red}{~\cite{Caprini:2015zlo}}. Interestingly, this frequency range overlaps with the frequency ranges which future space--based interferometers, such as LISA~\cite{Caprini:2015zlo}, DECIGO~\cite{Musha:2017usi}, and BBO~\cite{Crowder:2005nr}, will be most sensitive to. Discovering such a gravitational wave signal would establish
the EWPT as a new milestone in our understanding of the early universe. 

In the Standard Model (SM), electroweak symmetry breaking (EWSB) would proceed via a smooth crossover unless the Higgs mass is below $\sim 70$~GeV~\cite{Dine:1992vs,Kajantie:1995kf}. Therefore, the discovery of the SM Higgs boson with a mass $m_h = 125$~GeV~\cite{Aad:2012tfa,Chatrchyan:2012xdj} meant that the SM alone cannot satisfy the third Sakharov condition, i.e. departure from thermal equilibrium.~\footnote{Moreover, the amount of CP violation in the SM is too small for satisfying the second Sakharov condition, which means that new sources of CP violation should be considered. Addressing this problem is, however, beyond the scope of this work.} 

Consequently, the problem of strongly first order EWPTs was considered in several beyond the Standard Model (BSM) scenarios, most of them containing extra scalar states with respect to the SM. Examples include scalar singlet extensions~\cite{Noble:2007kk,Profumo:2007wc,Barger:2011vm,Wainwright:2012zn,Patel:2012pi,Profumo:2014opa,Katz:2014bha,Kozaczuk:2015owa,Chen:2017qcz,Vaskonen:2016yiu}, Two Higgs Doublet Models (2HDMs)~\cite{Bochkarev:1990fx,McLerran:1990zh,Turok:1990zg,Cohen:1991iu,Nelson:1991ab,Basler:2016obg}, the (Next to) Minimal Supersymmetric Standard Model~\footnote{In the MSSM, the 1st order EWPT is excluded by the measured Higgs mass, $m_h = 125$~GeV, and by light stop searches~\cite{Carena:2012np}.} (MSSM/NMSSM)~\cite{Menon:2004wv,Kozaczuk:2014kva}, or Composite Higgs Models~\cite{Creminelli:2001th,Randall:2006py,Nardini:2007me,Grinstein:2008qi,Espinosa:2011eu,Bruggisser:2018mus}. Scalars are of particular interest as, when integrating out heavy scalars, a $(H^{\dagger}H)^3$ term can be generated and therefore induce a barrier in the Higgs potential~\cite{Barger:2011vm,Huang:2015tdv}. Also, scalars contribute to a negative cubic term in the Higgs effective potential in the high temperature expansion and therefore can induce a barrier via thermal effects~\cite{Chung:2012vg}. There has been much less focus on fermionic extensions in the context of EW baryogenesis~\cite{Carena:2004ha,Davoudiasl:2012tu,Egana-Ugrinovic:2017jib,Baldes:2016gaf,Baldes:2016rqn,Braconi:2018gxo} as fermions do not contribute as much in the high temperature expansion. However, when the critical temperature is low compared to the new fermion masses, fermions contribute to the Higgs effective potential in the same way as scalars, and therefore can lead to non-trivial effects in the thermal history of the early universe. For example, Ref.~\cite{Egana-Ugrinovic:2017jib} considered two fermionic multiplets: a Majorana singlet and a vector--like (Dirac) $SU(2)_L$ doublet with the same quantum numbers as the SM leptonic doublet. In this setup, a strong first order EWPT can be induced by the neutral fermions, which couple strongly to the Higgs.

In this paper, we study a SM extension containing only Dirac fermions, and investigate the impact of these fermions on the thermal history of the Universe. We choose a BSM spectrum containing only vector--like leptons (VLLs), as current limits from the LHC push vector--like quarks (VLQs) at masses above $1$~TeV, making them too heavy to considerably influence the EWPT. We show that such a model can indeed accommodate a SFOPT capable of generating the BAU as long as the new fermions couple strongly to the Higgs. Interestingly, we find that, although having only one scalar field -- the SM Higgs, our model predicts a three--step phase transition, consisting of one smooth crossover at high temperatures and two SFOPTs at temperatures of the order of the EW scale. Moreover, we calculate the GW signal and collider impact of our model, comparing the signals with present and/or future collider (LHC) and GW searches. 

This paper is structured as follows. In Sec.~\ref{sec:new_fermions}, we construct a model containing new Vector-Like Leptons (VLLs), show how it can accommodate SFOPTs, and discuss the general predictions of this model. Sec.~\ref{sec:benchmarks} is dedicated to a more detailed analysis of three benchmark points, for which we calculate the GW signature and several collider observables, such as the VLL production cross sections and branching ratios. Finally, we summarize and present our conclusions in Sec.~\ref{sec:conclusions}.

\section{New Dirac Fermions and the Phase Structure of the Universe} \label{sec:new_fermions}

In this section, we build a minimal Dirac VLL model that can produce SFOPTs in the Early Universe. To calculate the PT strength in our model, we construct its 1-loop effective potential, comprising both zero-- and finite--temperature pieces, and then include the so-called Daisy resummation contribution. We then discuss the thermal behavior and current constraints of this model.

\subsection{A Minimal Vector-Like Lepton Model for Strong Phase Transitions}
\label{subsec:vll_model}

Let us start by outlining the logical steps followed when building our model. Firstly, we require that the new fermions couple strongly to the Higgs field. In the limit of null Yukawa couplings for the new fermions, adding them would leave the 1-loop effective potential unchanged with respect to the SM-only case. In order to dramatically change the thermal evolution of the 1-loop effective potential, which exhibits only a crossover in the SM, one should therefore consider strong Yukawas for the VLLs.

The simplest VLL model would correspond to adding a single VLL multiplet to the SM. In such a case, the only possible non-SM Yukawa terms would couple a SM lepton and the VLL multiplet to the Higgs doublet. However, such a term would mix the VLL and the SM lepton (which we assume to be the $\tau$ lepton, to avoid stronger constraints from electrons and muons). As explained in the previous paragraph, the new Yukawa coupling has to be large, which would result in a strong mixing between the VLL and $\tau$ and hence a significant departure from the SM prediction of the $\tau$ couplings. This forces us to discard such a scenario, as the $\tau$ couplings are tightly constrained to be SM-like by experimental measurements such as  $Z\to\tau\tau$ decays at LEP ~\cite{ALEPH:2005ab} or $h\to\tau\tau$ decays at LHC~\cite{CMS-PAS-HIG-17-031,Aaboud:2018pen}. Therefore, throughout this section, we neglect the mixing between VLLs and the SM fermions~\footnote{We will come back to this point in Sec.~\ref{sec:benchmarks}}, as the corresponding Yukawa couplings would have an insignificant effect on the phase structure of the Universe.

The next logical choice would be to augment the SM with one VLL doublet and one VLL singlet, since this configuration would allow for a Yukawa term coupling the two VLL multiplets to the Higgs doublet. However, such a model with strong Yukawas would badly violate custodial symmetry, giving rise to unacceptable contributions to the $T$ parameter~\cite{Peskin:1990zt,Peskin:1991sw,Altarelli:1990zd}. As we have checked, one cannot accommodate SFOPTs in this scenario without dramatically exceeding the experimental bounds on the $T$ parameter.

Therefore, the minimal solution is to add one VLL doublet and two VLL singlets, since such a configuration can accommodate an (approximate) custodial symmetry, which allows for large Yukawas while avoiding significant contributions to the $T$ parameter. We choose the new leptons to have similar $SU(3)_c \times SU(2)_L \times U(1)_Y$ quantum numbers as their SM counterparts:
\begin{equation}
 L_{L,R} = \left(\!\!\!\begin{array}{c}
 N \\  E 
\end{array}\!\!\!\right)_{L,R}   \sim (1, 2)_{-1/2}, \quad
N^\prime_{L,R} \sim  (1, 1)_0, \quad
 E^\prime_{L,R} \sim  (1, 1)_{-1},
\label{eq:vll_multiplets}
\end{equation}
where $L,R$ stand for the VLL chiralities. The new fermions $N^{(\prime)}_{L,R}$ and $E^{(\prime)}_{L,R}$ have zero and $-1$ electric charge, respectively, so we shall refer to them as neutral VLLs or VL neutrinos and charged VLLs or VL electrons. For denoting the multiplets in the equation above, we use the standard $(SU(3)_c, SU(2)_L)_Y$ notation, where the hypercharge $Y$ is given by the difference between the electric charge and the third isospin componen, i.e. $Y=Q-T_3$.

As stated previously, we neglect for the time being the mixing between the SM leptons and the VLLs. Therefore, the most general renormalizable VLL Yukawa Lagrangian, consistent with $SU(3)_c \times SU(2)_L \times U(1)_Y$ gauge symmetry, reads
\begin{align}
-{\cal L}_{VLL}&= y_{N_R} \overline{L}_L \tilde{H} N_R^\prime + y_{N_L} \overline{N}^\prime_L \tilde{H}^\dagger L_R 
 +y_{E_R} \overline{L}_L H E^\prime_R + y_{E_L} \overline{E}^\prime_L H^\dagger L_R  \notag \\
&+m_L \overline{L}_L L_R 
+ m_N \overline{N}^\prime_L N^\prime_R+m_E \overline{E}^\prime_L E^\prime_R + \mathrm{h.c.} \; ,
\label{eq:vll_yuk_lagrangian}
\end{align}
where $H$ represents the SM Higgs doublet and $\tilde{H}$ its $SU(2)$ conjugate, $y$'s the dimensionless Yukawa couplings, and $m$'s the vector-like masses. Upon Electroweak Symmetry Breaking (EWSB), one can write the neutral and charged mass matrices, respectively, as
\begin{equation}
\mathcal{M}_N = \begin{pmatrix}
m_L & v_h \, y_{N_L} \\ v_h \, y_{N_R} & m_N
\end{pmatrix}, \quad 
\mathcal{M}_E = \begin{pmatrix}
m_L & v_h \, y_{E_L} \\ v_h \, y_{E_R} & m_E
\end{pmatrix},
\label{eq:vll_mass_matrices}
\end{equation}
with $v_h \equiv v/\sqrt{2} \simeq 174$~GeV, where $v \simeq 246$~GeV is the Higgs vacuum expectation value (vev). The physical masses, which we denote and order as $m_{N_1} < m_{N_2}$ and $m_{E_1} < m_{E_2}$, are obtained as usual by bi-diagonalizing the mass matrices from the above equation. The physical couplings of the VLL eigenstates can be calculated from the corresponding rotation matrices that bi-diagonalize the mass matrices.

\subsection{Phase Transition Calculation}
\label{subsec:pt_calc}

To study the thermal history of the Universe, we first write down the 1-loop effective scalar potential, taking into account the effect of SM particles strongly coupled to the Higgs ($W$ and $Z$ bosons, $t$ quark, $h$ boson, and Goldstone bosons, $\chi$) and of the VLLs. We denote the background field-dependent squared mass as $\omega_i (\phi)$, where $i$ labels the particles and $\phi$ is the background field value.
For the SM fields coupling strongly to the Higgs, the various $\omega$'s are
\begin{gather}
\omega_{W,Z} (\phi) = \frac{m_{W,Z}^2}{v^2} \phi^2, \quad \omega_t (\phi) = \frac{m_{t}^2}{v^2} \phi^2, \notag \\
\omega_h (\phi) = \frac{m_h^2}{2 v^2} (3 \phi^2 - v^2), \quad \omega_{\chi} (\phi) = \frac{m_h^2}{2 v^2} (\phi^2 - v^2).
\label{eq:SM_field_dep_masses}
\end{gather}
Throughout this work, we use the following values for the masses of the SM particles~\cite{Tanabashi:2018oca}:
\begin{equation}
m_W = 80.4 \, {\rm GeV}, \: m_Z = 91.2 \, {\rm GeV}, \: m_t = 174 \, {\rm GeV}, \: m_h = 125 \, {\rm GeV}.
\label{eq:SM_masses}
\end{equation}
The $\phi$-dependent VLL squared eigenmasses are obtained by diagonalizing $\mathcal{M}_X^{\dagger} \mathcal{M}_X$, with $X=N,E$, and $\mathcal{M}_X$ defined in Eq.~\eqref{eq:vll_mass_matrices}. We thus have:
\begin{gather}
\omega_{X_{1,2}}  (\phi) = \frac{1}{2} \Bigg( m_L^2 + m_X^2 + \frac{y_{X_L}^2 + y_{X_R}^2}{2} \phi^2  \notag \\  \mp \sqrt{ \left( m_L^2 + m_X^2 + \frac{y_{X_L}^2 + y_{X_R}^2}{2} \phi^2 \right)^2 - \left( 2 m_L m_X - y_{X_L} y_{X_R} \phi^2 \right)^2} \Bigg),
\end{gather}
where ``$-$'' corresponds to $X_1$ (the lighter eigenstate) and ``$+$'' to $X_2$ (the heavier eigenstate). To make the connection with previous notations, we remind the reader that $\omega_{X_j} (v) = m_{X_j}^2$, with $j=1,2$.

\subsubsection*{The Effective Potential}

We now proceed to calculating the 1-loop effective potential, comprised of the SM tree level part and the zero and finite temperature 1--loop contributions (for both the SM particles and the VLLs), to which we add the so-called Daisy contribution. The Daisy (or ring) contribution is a finite temperature effect coming from the resummation of higher--loop IR-divergent diagrams of a certain topology~\cite{Arnold:1992rz}, whose sum amounts to a finite result. 

\textbf{The SM tree level contribution} is given by
\begin{equation}
V_0 (\phi) = \frac{m_h^2}{8 v^2} (\phi^2 - v^2)^2.
\label{eq:VSM_tree}
\end{equation}
%From a formal point of view, the tree-level potential should also include counter-terms for the constant, quadratic, and quartic terms (in $\phi$), which would cancel the ultraviolet (UV) divergences coming from the 1-loop contribution to the potential. Instead of writing the counter-terms explicitly, we choose to express in what follows only the finite part that remains after the cancellation of infinities from the 1-loop contribution against the counter-terms.

\textbf{The 1-loop zero-temperature contribution.}
 %TK Cut-off regularization In order to obtain a finite piece for the zero-temperature contribution arising at 1-loop, renormalization conditions need to be specified, which in turn fix the expressions for the counter-terms, allowing for a cancellation of infinities. Following Ref.~\ref{Quiros_intro}, 
We work in the on-shell renormalization scheme for the 1-loop contribution,
%and impose through the renormalization conditions that the position of the EW minimum and the value of the Higgs mass remain the same as their tree-level counterparts. Mathematically, these two conditions read:
\begin{equation}
\frac{{\rm d} \Delta V_1^0}{{\rm d} \phi}\bigg|_{\phi = v} = \frac{{\rm d}^2 \Delta V_1^0}{{\rm d} \phi^2}\bigg|_{\phi = v} = 0,
\label{eq:renorm_cdts}
\end{equation}
where $\Delta V_1^0 (\phi)$ is the (finite) Coleman-Weinberg potential, which includes the counter-terms and the zero-temperature
% (or temperature-independent) 
 piece of the 1-loop correction to the potential. We further split $\Delta V_1^0 (\phi)$ into two pieces, 
\begin{equation}
\Delta V_1^0 (\phi) = \Delta V_{\rm 1, \, SM}^0 (\phi) + \Delta V_{\rm 1, \, VLL}^0 (\phi),
\label{eq:1-loop_SM-VLL_split}
\end{equation}
$\Delta V_{\rm 1, \, SM}^0$ comprising the effect of SM particles and $\Delta V_{\rm 1, \, VLL}^0$ capturing the VLL contribution (at $T=0$). In the Landau gauge, which we are going to use throughout this work, these two contributions read~\footnote{Note that, since $\omega_\chi(v) = 0$, the Goldstone contribution at 1--loop diverges due to the logarithmic term. This IR divergence can be cured by imposing that the second derivative of the total potential (tree plus 1--loop contributions)  at $\phi = v$ is equal to the 1--loop Higgs mass evaluated at $p^2 = m_h^2$ (and not $p^2 = 0$), where $p$ is the external momentum in the Higgs self-energy diagram. This amounts to replacing $\omega_\chi(v)$ with $\omega_h(v)$~\cite{Anderson:1991zb} in the logarithmic term corresponding to $i=\chi$ from Eq.~\eqref{eq:V10_SM}. For a more detailed discussion of the matter, see Refs.~\cite{Espinosa:1992kf,Delaunay:2007wb}.}
\begin{align}
\Delta V_{\rm 1, \, SM}^0 (\phi) &= \sum_{i=W,Z,h,\chi,t} \frac{n_i}{64 \pi^2} \left[ \omega_i^2 (\phi) \left( \log \frac{\omega_i(\phi)}{\omega_i(v)} - \frac{3}{2} \right) + 2 \,  \omega_i(v) \omega_i(\phi) \right], \label{eq:V10_SM} \\
\Delta V_{\rm 1, \, VLL}^0 (\phi) &= \sum_{i =  N_{1,2},E_{1,2}} \frac{n_{\rm VLL}}{64 \pi^2} \left[ \omega_i^2 (\phi) \left( \log \frac{\omega_i(\phi)}{\mu^2} - \frac{1}{2} \right) + a_i \phi^2 - \frac{b_i}{v^2} \phi^4 \right], \label{eq:V10_VLL}
\end{align}
where the coefficients $a_i$ and $b_i$, with $i=N_1,N_2,E_1,E_2$, are given by~\cite{Carena:2004ha,Egana-Ugrinovic:2017jib}
\begin{align}
a_i &= \frac{1}{2} \left[ \left( \omega_i^{\prime 2} + \omega_i \omega_i^{\prime\prime} - 3 \frac{\omega_i \omega_i^{\prime}}{v} \right) \log \frac{\omega_i}{\mu^2} + \omega_i^{\prime 2} \right], \label{eq:ai_quadratic_phi}\\
b_i &= \frac{1}{4} \left[ \left( \omega_i^{\prime 2} + \omega_i \omega_i^{\prime\prime} - \,\,\, \frac{\omega_i \omega_i^{\prime}}{v} \right) \log \frac{\omega_i}{\mu^2} + \omega_i^{\prime 2} \right], \label{eq:bi_quartic_phi}
\end{align}
In Eqs.~\eqref{eq:V10_VLL}--\eqref{eq:bi_quartic_phi}, $\mu$ denotes an arbitrary energy scale which is introduced to make the logarithm arguments adimensional. Moreover, we used the following notations:
\begin{equation}
\omega_i \equiv \omega_i (v), \qquad \omega_i^{\prime} \equiv \frac{{\rm d} \omega(\phi) }{{\rm d} \phi}\bigg|_{\phi = v}, \quad {\rm and} \quad \omega_i^{\prime\prime} \equiv \frac{{\rm d}^2 \omega(\phi) }{{\rm d} \phi^2}\bigg|_{\phi = v}. \label{eq:omega_prime}
\end{equation}
Finally, the number of degrees of freedom for the fields running in the loops are 
\begin{equation}
n_{\{W,Z,h,\chi,t,{\rm VLL}\}} = \{6,3,1,3,-12,-4\}, \label{eq:dofs}
\end{equation}
with $n_{\rm VLL} = n_{N_{1,2}} = n_{E_{1,2}} = -4$, since the VLLs we introduce are Dirac fermions.

\textbf{The 1-loop finite-temperature contribution} is obtained, in the imaginary time formalism, by performing a Wick rotation and compactifying the resulting Euclidean time dimension on a circle of radius $R=(2\pi T)^{-1}$, with $T$ denoting the temperature. The fields are thus Fourier expanded along the periodic time dimension, with eigenfrequencies given by the Matsubara frequencies, which are given by $2 n \pi T $ for bosons (periodic on the time circle) and $(2 n + 1) \pi T $ for fermions (anti--periodic on the time circle). Performing the infinite sum on these frequencies gives rise to the finite--$T$ contribution to the potential, which reads
\begin{equation}
\Delta V_1^T (\phi, T) = \frac{T^4}{2 \pi^2} \left[ \sum_{i=W,Z,h,\chi} n_i J_b \left( \frac{\omega_i(\phi)}{T^2} \right) + \sum_{i=t,N_{1,2},E_{1,2}} n_i J_f \left( \frac{\omega_i(\phi)}{T^2} \right) \right]. \label{eq:V1T}
\end{equation}
Here, $J_b$ and $J_f$ are adimensional integrals accounting for the thermal contribution of boson and fermions, respectively, and are given by (see e.g. Ref.~\cite{Quiros:1994dr})
\begin{equation}
J_{b,f} (x^2) = \int_0^{\infty} {\rm d}k \, k^2 \log \left[ 1 \mp {\rm e}^{-\sqrt{k^2 + x^2}} \right],
\label{eq:thermal_integrals}
\end{equation}
where the ``$-$'' (``$+$'') sign applies to bosons (fermions).

\textbf{The Daisy contribution.}
At finite temperature, the perturbative expansion in terms of a small coupling breaks down due to IR divergences coming from thermal loops involving bosonic 0--modes, which have vanishing Matsubara frequencies. This problem can be fixed by performing a resummation of the a certain class of diagrams, the so--called Ring or Daisy diagrams, which are $N$--loop diagrams in which $N-1$ loops are each attached to the main one through one and only one 4--point vertex. While all these diagrams are IR--divergent if taken one by one, their sum adds up to a finite result, which corresponds to the following contribution to the effective potential:
\begin{equation}
\Delta V_{\rm D} (\phi, T) = \frac{T}{12\pi} \sum_{i=h,\chi,W,Z,\gamma} \bar{n}_i \left[ \omega_i^{3/2} (\phi) - \left( \omega_i (\phi) + \Pi_i (T) \right)^{3/2} \right],
\label{eq:V_daisy}
\end{equation}
with $\bar{n}_{\{h,\chi,W,Z,\gamma\}} = \{1,3,2,1,1\}$ represent  either scalar ($h,\chi$) or gauge boson longitudinal degrees of freedom, since only these degrees of freedom acquire a thermal mass (transverse gauge modes are protected by gauge symmetry). Note that fermionic diagrams do not produce any IR divergences, as their 0--modes have non-vanishing Matsubara frequencies. In  the equation above, the $\Pi_i$'s represent the thermal or Debye masses of the fields appearing in the sum
% ,for which we used formulas adapted from Ref.
~\cite{Delaunay:2007wb},
\begin{align}
\Pi_{h,\chi} (T) &= \frac{T^2}{4 v^2} \left( m_h^2 + m_Z^2 + 2 m_W^2 + 2 m_t^2 \right), \label{eq:pi_scalars} \\
\Pi_W (T) &= \frac{22 \, T^2}{3 v^2} m_W^2 , \label{eq:pi_W} \\
\Pi_Z (T) &= \frac{22 \, T^2}{3 v^2} \left(m_Z^2 - m_W^2 \right) - \omega_W (\phi), \label{eq:pi_Z} \\
\Pi_\gamma (T) &= \omega_W (\phi) + \frac{22 \, T^2}{3 v^2} m_W^2. \label{eq:pi_gamma}
\end{align}
In the expressions above, we use the high--temperature approximation for the contribution of the SM particles, which is justified by the fact that the Daisy diagrams give a negligible contribution for $\omega_i (\phi) \gtrsim T^2$. Moreover, because the strong phase transitions in our model occur at temperatures well below the VLL masses, we neglect the VLL contribution to the $\Pi_i$'s in Eqs.~\eqref{eq:pi_scalars}--\eqref{eq:pi_gamma}.

Adding all the pieces together, the effective potential we use for calculating the strength of the phase transition is given by
\begin{equation}
V(\phi,T) = V_0 (\phi) + \Delta V^0_{\rm 1, SM} (\phi)+ \Delta V^0_{\rm 1, VLL} (\phi) + \Delta V^T_1 (\phi, T) + \Delta V_D (\phi, T),
\label{eq:V_full}
\end{equation}
with all the contributions detailed in Eqs.~\eqref{eq:VSM_tree}, \eqref{eq:V10_SM}, \eqref{eq:V10_VLL}, \eqref{eq:V1T}, and \eqref{eq:V_daisy}. Finally, since only potential differences have a physical impact in our analysis, we shift the potential by a constant such that $V(\phi=0, T) = 0$ for every $T$.  

\subsubsection*{Scan for the PT strength calculation }

We now calculate the strength of the phase transition in our model. We scan over the following range of parameter values:
\begin{equation}
m_L, m_N, m_E \in [500,1500]~{\rm GeV}, \qquad y_{N_{L,R}}, y_{E_{L,R}} \in [2, \sqrt{4\pi}].
\end{equation}
In our initial scans, we allowed for wider ranges, and found out that parameter values inside the ranges shown above are more likely to lead to strong phase transitions. Also, as noted in Ref.~\cite{Egana-Ugrinovic:2017jib}, having $y_{N_R} y_{N_L} > 0$ and $y_{E_R} y_{E_L} > 0$ favors SFOPTs, which is why we chose all the Yukawas to be positive.

After each point in parameter space is generated, we check whether the said point is in agreement with experimental constraints. Firstly, as the VLLs under consideration have $SU(2)_L$ quantum numbers, they affect the electroweak gauge boson self-energies, so one constraint is the contribution to the oblique parameters $S$ and $T$~\cite{Peskin:1991sw,Peskin:1990zt}, for which we use the $2\sigma$ values quoted in Ref.~\cite{Baak:2014ora}. Secondly, the charged VLLs change the loop-induced $h\gamma\gamma$ coupling with respect to its SM value. This coupling is probed at the LHC through the diphoton Higgs signal strength, $\mu_{\gamma\gamma} \equiv \Gamma(h\to\gamma\gamma) / \Gamma(h\to\gamma\gamma)_{\rm SM}$. As the experimental bound for this observable, we use the 2$\sigma$ interval quoted by the ATLAS collaboration%in Ref.
, $0.71 < \mu_{\gamma\gamma} < 1.29$~\cite{Aaboud:2018xdt}.~\footnote{A subsequent CMS measurement shows a slightly higher value for the $h\to\gamma\gamma$ signal strength, $\mu_{\gamma\gamma} = 1.18^{+0.17}_{-0.12}$ at 1$\sigma$~\cite{Sirunyan:2018ouh}.} Thirdly, we impose a lower bound on the masses of the lighter eigenstates, $ m_{E_1} > 100$~GeV and $m_{N_1} > 90$~GeV~\cite{ALEPH:2005ab}.

From the theoretical point of view, we also impose a lower limit on the depth of the EW minimum at the lower minimum, $\left |V(\phi=v,T=0) \right|$ (we remind the reader that, by convention, $V(0,T)=0$). As illustrated in Ref.~\cite{Harman:2015gif},  the lower the depth of the present day EW minimum, the more delayed and thus the stronger the phase transition is. For our analysis, we choose $\left |V(\phi=v,T=0) \right| < 8.5 \times 10^{7}$~GeV$^4$, a value for which we checked explicitly that most of the surviving points exhibit SFOPTs.

For the points surviving the constraints listed above, we calculate the phase transition strength (or order parameter), which is defined as $\xi \equiv \phi_c / T_c$, where $\phi_c$ and $T_c$ are the critical field value and critical temperature, respectively. $T_c$ is defined as the temperature at which the values of the potential at the minima located at $\phi=0$ (``symmetric minimum'') at $\phi \neq 0$ (``broken minimum'') become degenerate. $\phi_c$ is defined as the position along the field axis of the broken minimum for $T=T_c$:
\begin{equation}
V(\phi_c,T_c)=0 \qquad {\rm and} \qquad V^{\prime} (\phi_c,T_c)=0 \qquad {\rm with} \quad \phi_c \neq 0 \,.\label{eq:phic_Tc_eqs}
\end{equation} 
In order to find the critical field value and temperature for a given point in the parameter space spanned by the VLL Yukawas and diagonal mass terms, we numerically solve the system of equations Eq.~\eqref{eq:phic_Tc_eqs}, whose solution is given by $\phi_c$ and $T_c$ from which we calculate the phase transition strength, $\xi = \phi_c/T_c$. In the above equations, the prime symbol denotes differentiation with respect to $\phi$.  

To speed up computations, without spoiling the reliability of our results, we calculate $\phi_c$ and $T_c$ for the 1-loop effective potential without the subleading scalar (Higgs+Goldstone) and Daisy contributions. We checked that adding taking into account these contributions amounts to a $\mathcal{O}(5-10\%)$ correction to the values of $\xi$, which justifies \textit{a posteriori} our approximation.~\footnote{Nevertheless, when studying the benchmark points in Sec.~\ref{sec:benchmarks}, we are going to take into account the scalar and Daisy contributions too.} Therefore, in the following, unless mentioned otherwise, all the scatter plots resulting from our scan will not feature the scalar and Daisy contributions.

\subsection{General Predictions of the Model}
\label{subsec:gen_predictions}

\begin{figure}[t!]
  \centering
  \includegraphics[width=0.95\textwidth]{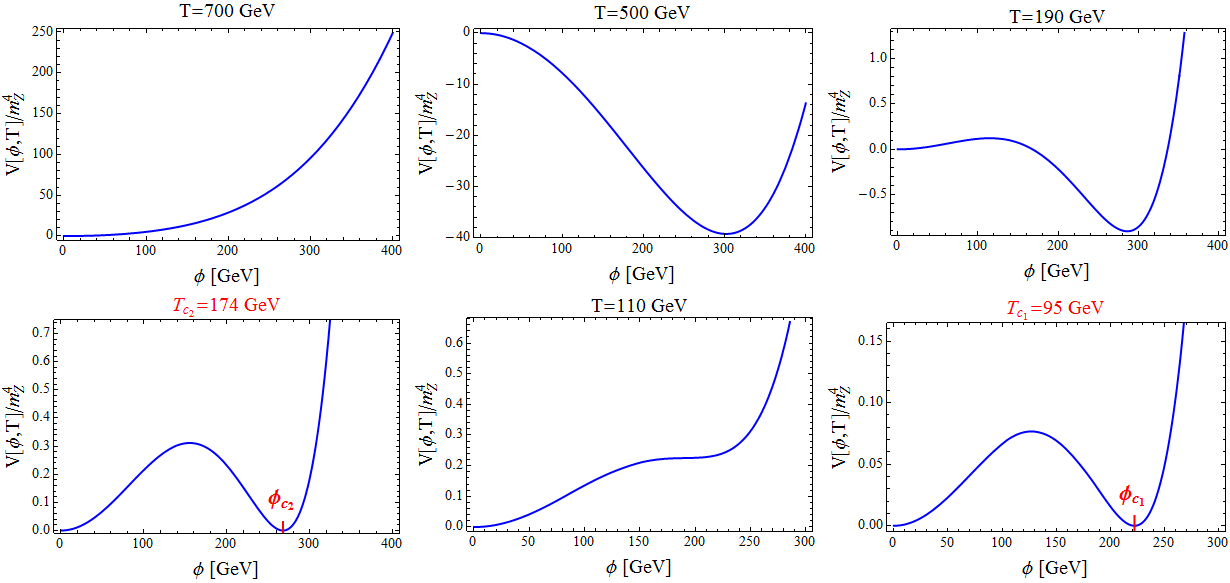}
  \caption{\small Thermal evolution of the potential, with $V(\phi,T)$ (in units of $m_Z^4$) plotted as a function of the background field value, $\phi$, for six temperature values: $T=700$, $500$, $190$, $174$, $110$, and $95$~GeV. In these plots, the scalar and Daisy contributions have been taken into account when calculating the effective potential.}
  \label{fig:V-thermal-evolution}
\end{figure}

We now present some general predictions of our model. The most striking one concerns the non-trivial thermal evolution of the potential from Eq.~\eqref{eq:V_full}. For the scan points that pass the constraints, three distinct phase transitions are predicted in the Early Universe: one crossover and two SFOPTs. As an illustration of this fact, we show in Fig.~\ref{fig:V-thermal-evolution} the behavior of the potential for the benchmark point BM1 presented in Sec.~\ref{sec:benchmarks}. In Fig.~\ref{fig:V-thermal-evolution}, we plot the potential as a function of the background field value, $\phi$, for six different temperatures. In its early stages, the Universe starts in the symmetric phase, i.e. at $\phi=0$, and then, as it expands and cools down, has a crossover to the broken phase, $\phi \neq 0$, which results in EWSB. In the following, we shall refer to the broken phase also as the EW or broken minimum. As the temperature lowers, the EW minimum becomes less and less deep, and a potential barrier starts developing between the symmetric and broken phases. At a critical temperature, which we denote by $T_{c_2}$, the two minima become degenerate, and the Universe starts tunneling back to the symmetric phase, thus undergoing a SFOPT and restoring EW symmetry. We denote the critical field value for this first SFOPT as $\phi_{c_2}$, and the corresponding PT strength as $\xi_2 = \phi_{c_2} / T_{c_2}$. After this SFOPT, as the Universe cools down, the EW minimum becomes a local minimum, and for this particular point in parameter space even disappears altogether. However, this trend reverses at even lower temperatures, and the EW minimum becomes again degenerate with the symmetric one, and EW symmetry breaking is triggered once again. This happens for a critical temperature $T_{c_1}$ and a corresponding critical field value $\phi_{c_1}$, with the PT strength given by $\xi_1 = \phi_{c_1} / T_{c_1}$. For temperatures below $T_{c_1}$, the Universe remains in the EW minimum.

%%%%%%%%%%%%%%%%%%
%%%%%%%%%%%%%%%%%%

We now discuss the non--trivial behavior of the potential, as shown in Fig.~\ref{fig:V-thermal-evolution}, focusing on the contributions determining the three PTs -- the crossover and the two SFOPT. Firstly, the symmetry--breaking crossover occurring at high $T \gtrsim 500-600$~GeV is induced through thermal effects: the light VLL ($N_1$ and $E_1$) thermal contribution, which favors the broken phase, competes with the SM and heavy VLL ($N_2$ and $E_2$) thermal pieces, which tend to keep the Universe in the symmetric phase. The crossover takes place when the first contribution overtakes the other one. Secondly, the earlier, symmetry--restoring SFOPT, which typically takes place at $T \sim 150-200$~GeV, is sourced by both finite and zero temperature effects. In this case, the SM thermal part and the $T=0$ VLL contribution tend to restore EW symmetry, whereas the light VLL thermal part and the SM $T=0$ contribution work towards keeping the Universe in the broken minimum. The net effect is a barrier between the two minima, and the symmetry--restoring PT is thus strongly first--order. Concerning the heavy VLLs, they are decoupled from the thermal bath at this stage and play a negligible role in the dynamics of the earlier SFOPT. Finally, the later SFOPT, which occurs at temperatures around $100$~GeV and is responsible with generating the BAU, is again the result of an interplay between zero-- and finite--T effects. On one side, the SM thermal piece and the $T=0$ VLL contribution favor the Universe to remain in the symmetric phase, while the SM $T=0$ part favors EWSB. Adding together these competing effects, a barrier is again generated between the two minima, and EW symmetry is broken via a SFOPT. The thermal impact of the new fermions is negligible for this PT, as both light and heavy VLLs are decoupled from the plasma when this PT occurs.

Among the two SFOPTs in this model, only the later one is responsible for generating the BAU. During the later one, the universe undergoes a phase transition from the symmetric phase to the broken phase. This leads to the formation and expansion of bubbles of the true vacuum in the false, symmetric vacuum. In the presence of the $CP$ violation, particle interactions with the expanding bubbles can lead to the creation of an excess of baryons inside the bubbles through baryon number violating processes induced by sphalerons~\cite{PhysRevD.30.2212}. The sphaleron process, when in equilibrium, wipes off the created excess of baryons. The sphaleron rate in the broken phase is proportional to ${\rm e}^{-\sqrt{\omega_W(\phi)}/T}$, and is suppressed if the phase transition is of strong first order, $\xi$, is greater than $1.3$~\cite{SHAPOSHNIKOV1987757}. In our case, this condition translates to $\xi_1 \gtrsim 1.3$, which, as we are going to see in the following, is satisfied by most of the surviving points from our scan. The earlier phase transition, however, can not generate baryon asymmetry. The sphaleron rate is unsuppressed and behaves as $T^4$~\cite{SHAPOSHNIKOV1987757} in the symmetric phase, so any matter anti--matter asymmetry thus generated would be wiped out by the high sphaleron rate in the symmetric phase, to which the Universe tunnels during the earlier SFOPT.

%%%%%%%%%%%%%%%%%%
\begin{figure}[t!]
  \centering
  \includegraphics[width=0.6\textwidth]{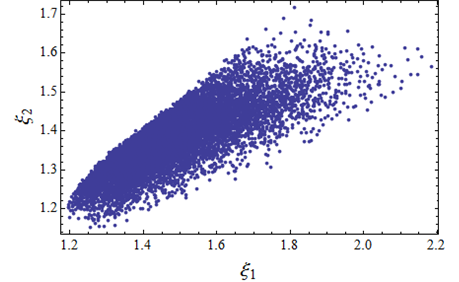}
  \caption{\small Strength of the earlier phase transition, $\xi_2$, plotted against the strength of the late phase transition, $\xi_1$.  }
  \label{fig:xi1_xi2}
\end{figure}
%%%%%%%%%%%%%%%%%%

We now display some scatter plots resulting from our scan. As mentioned previously, the points shown in these scatter plots correspond to a computation done without incorporating the scalar and Daisy contributions to the effective potential.

%%%%%%%%%%%%%%%%%%
\begin{figure}[t!]
  \centering
  \includegraphics[width=0.45\textwidth]{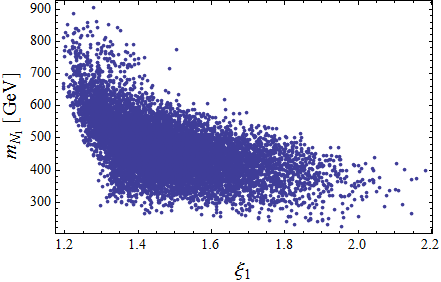} \qquad \includegraphics[width=0.45\textwidth]{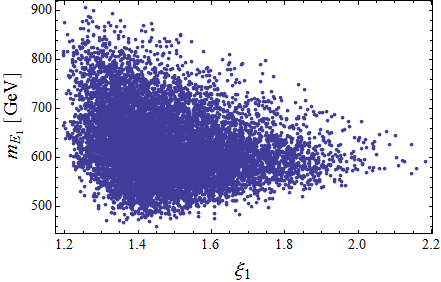}
  \caption{\small Correlation between the strength of the later phase transition, $\xi$, and the mass in GeV of the lighter neutral (left panel) and charged (right panel) VLL eigenstate.}
  \label{fig:xi1_masses}
\end{figure}
%%%%%%%%%%%%%%%%%%

Firstly, we display in Fig.~\ref{fig:xi1_xi2} the correlation between $\xi_1$ and $\xi_2$, the strengths of the two SFOPTS. A feature implied by this plot is that, in our model, the later PT is in general stronger than the earlier one, i.e. $\xi_1 > \xi_2$. In simple terms, this can be explained by noting that the critical field values are rather close, $\phi_{c_1} \simeq \phi_{c_2} \simeq \mathcal{O}(v)$ (see the example in Fig.~\ref{fig:V-thermal-evolution}), whereas there is a larger separation between the two critical temperatures, $T_{c_2} > T_{c_1}$. Furthermore, as we shall see in Sec.~\ref{sec:benchmarks}, the later SFOPT generally produces a stronger Gravitational Wave (GW) signal than the earlier one, as $\xi_1 > \xi_2$.

Secondly, we plot in Fig.~\ref{fig:xi1_masses} the masses of the lighter VLL eigenstates, $m_{N_1}$ and $m_{E_1}$, versus the corresponding values of $\xi_1$. We find that, when applying the constraints mentioned in Sec.~\ref{subsec:pt_calc}, the lightest new particles predicted by our model lie in the hundreds of GeV range, with masses typically comprised between $250$~GeV and $900$~GeV. We also notice a correlation between the masses of $N_1$, $E_1$ and the strength of the later PT: the larger is $\xi_1$, the lower are the upper bounds of $m_{N_1}$ and $m_{E_1}$. This correlation can be understood by using a decoupling argument: the more massive the new fermions become, the more we expect their influence on EW scale physics to diminish. It is also worth noting that, in the right panel of Fig.~\ref{fig:xi1_masses}, as $\xi_1$ increases, the lower bound on the mass of $E_1$ increases. This behavior is induced by the $\mu_{\gamma\gamma}$ constraint: larger values of $\xi_1$ need larger values of Yukawa couplings for the charged VLLs, which, in order to satisfy the limit on $\mu_{\gamma\gamma}$, need to be compensated by a larger $m_{E_1}$.

\begin{figure}[t!]
  \centering
  \includegraphics[width=0.6\textwidth]{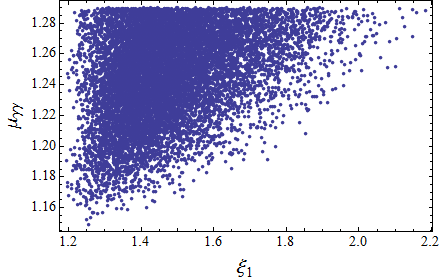}
  \caption{\small Scatter plot of the strength of the later phase transition, $\xi$, versus the $h\to\gamma\gamma$ signal strength, $\mu_{\gamma\gamma}$.}
  \label{fig:xi1_digamma}
\end{figure}

Finally, we show in Fig.~\ref{fig:xi1_digamma} the correlation between the Higgs-diphoton signal strength, $\mu_{\gamma\gamma}$, and $\xi_1$. The diphoton signal strength is influenced in our model by the charged VLLs running in the $h\to\gamma\gamma$ loop. We observe a clear trend in the figure: the stronger the later SFOPT is, the more $\mu_{\gamma\gamma}$ deviates from its SM value, $\mu_{\gamma\gamma}^{\rm SM} = 1$. Moreover, the deviation of the diphoton signal strength is always positive, our model predicting an increase of order 15--30$\%$ in $\mu_{\gamma\gamma}$. This correlation can be understood as follows: in order to obtain strong PTs, the new VLLs -- both neutral and charged -- need to be not too heavy and to have large Yukawa couplings to the SM Higgs. In turn, relatively light charged VLLs which couple strongly to the Higgs give a sizeable contribution to the $h\gamma\gamma$ loop, hence the deviation in $\mu_{\gamma\gamma}$. The positive interference between the charged VLL and SM contributions comes from the fact that strong SFOPTs favor same--sign Yukawa couplings in the charged mass matrix written in Eq.~\eqref{eq:vll_mass_matrices}. We also mention that $\xi_1$ and the $S,T$ parameters are uncorrelated.

%%%%%%%%%%%%%%%%%%
%%%%%%%%%%%%%%%%%%

Before moving to the next section, we would like to stress the importance of  $\mu_{\gamma\gamma}$ in testing our model. As explained in the previous paragraph, for our model, achieving a FOPT strong enough to accommodate the generation of the BAU entails a deviation of more than 10$\%$ of the loop-induced $h\gamma\gamma$ coupling from its SM value. The $\mu_{\gamma\gamma}$ observable is currently measured by the two LHC collaborations, ATLAS and CMS, and the present error is of order $\sim 15\%$ at $1\sigma$~\cite{Aaboud:2018xdt,Sirunyan:2018koj}. However, in the future high--luminosity phase of LHC, HL--LHC, the error $\mu_{\gamma\gamma}$ is expected to reduce to $\sim 5\%$~\cite{ATLAS:2013hta,CMS:2013xfa,ATL-PHYS-PUB-2014-016}. Therefore, we conclude that the HL--LHC will be able to fully test our model of VLL--induced EW baryogenesis.

\section{Gravitational wave and collider signatures}
\label{sec:benchmarks}

In this section, we perform an in-depth analysis of three benchmark points selected from our model. The first benchmark point corresponds to the strongest later first-order PT in the parameter space. The second benchmark point has a relatively lower value of $\mu_{\gamma\gamma}$, but at the same time accommodating two SFOPTs. The third benchmark point corresponds to the lowest value (among the points retained from our scan) of the mass of the lightest charged eigenstate, $m_{E_1}$. For all three benchmark points, the later PT, which is the one responsible for generating the BAU, is strongly first order, $\xi_1 > 1.3$. Moreover, the Higgs diphoton signal strength deviates from unity, as expected from the high Yukawas, but is still within the $2\sigma$ interval quoted by the most recent ATLAS measurement~\cite{Aaboud:2018xdt}. 

Furthermore, the masses of the lightest VLLs are in the $400$--$800$~GeV range, and the (correlated) values of the $S$ and $T$ parameters are well within the $2\sigma$ range. The benchmark scenarios are listed in Table~\ref{tab:table-bm}. In addition to the observables discussed in Sec.~\ref{subsec:gen_predictions}, we calculate the gravitational wave (GW) spectrum and several collider observables (VLL branching ratios and production cross sections) corresponding to these benchmarks and comment on the detectability of such signals at future GW experiments and the LHC. We remind that in this section we take into account the scalar and Daisy contributions as well when calculating the effective potential.

\begin{table}[h]
\renewcommand{\arraystretch}{1.6}
\centering
\begin{tabular}{|c|c|c|c|}
\hline 
&BM1 & BM2 & BM3 \\ \hline
$y_{N_L}$& 3.40 & 3.47&3.47 \\ \hline
$y_{N_R}$ & 3.49 &3.45 & 3.36\\ \hline
$y_{E_L}$ & 3.34 & 3.33&2.55 \\ \hline
$y_{E_R}$ & 3.46 & 3.41&3.28 \\ \hline
$m_{L}$ (TeV)& 1.06 & 1.42& 1.43\\ \hline
$m_{N}$ (TeV) & 0.94 &0.75 & 0.83\\ \hline
$m_E$ (TeV) & 1.34 & 1.25& 0.72\\ \hline
$\xi_1$ & 2.34 &2 & 1.56\\ \hline
$\xi_2$ & 1.54 &1.35 & 1.38\\ \hline
$\mu_{\gamma\gamma}$ &1.28 & 1.20& 1.28\\ \hline
 $\Delta \chi^2 (S,T)$ & 1.33 & 3.60&4.57 \\ \hline
 $m_{N_1}$ (GeV) & 400 &401 &466 \\ \hline
 $m_{E_1}$ (GeV) & 592 &740 & 460\\ \hline
\end{tabular}
\caption{\small Benchmark Points.}
\label{tab:table-bm}
\end{table}	 

\subsection{GW signature}
\label{subsec:gw-signature}
A first order EWPT is expected to generate a stochastic background of gravitational waves~\cite{Hogan:1986qda,Kamionkowski:1993fg}. In a radiation--dominated Universe, there are three sources of GW production at a SFOPT:  bubble collisions, in which the localized energy density generates a quadrupole contribution to the stress-energy tensor, which in turn gives rise to gravitational waves, plus sound waves in the plasma and  magnetohydrodynamic turbulence. The latter two  are generated after the bubbles have collided.

There are two key parameters that determine the spectrum of the stochastic GW background generated during a SFOPT. The first one, denoted by $\alpha$, represents the ratio between the latent heat released during the PT and the radiation energy density at the temperature at which the PT occurs. The $\alpha$ parameter plays a role in setting the strength of the GW signal: the higher is $\alpha$, the stronger is the predicted GW stochastic background. The second one, commonly referred to as $\beta$, sets the inverse timescale associated with the PT duration. $\beta$ influences both the strength of the GW signal, as well as the frequency at which the GW peaks. A higher value of $\beta$ implies a weaker GW signal and a shifting towards higher values of the peak frequency of the signal. For details regarding our computation of the GW spectrum we refer the reader to the Appendix.

For the points of our scan that accommodate a strong enough first--order later PT, the typical values of these parameters are
\begin{equation}
\alpha \sim 0.01 - 0.1, \qquad \beta/H_{\rm PT} \sim 10^3 - 10^4. 
\end{equation}  
By comparison, the more popular singlet scalar models feature higher values of $\alpha$ and lower values of $\beta$~\cite{Huang:2016cjm,Marzola:2017jzl}, which is why our VLL model predicts weaker GW signals than the singlet extension of the SM. Moreover, due to the high values of $\beta/H_{\rm PT}$, the typical GW signal of our VLL model peaks at frequencies in the $0.01 - 1$~Hz range. This can be understood immediately from Eq.~\eqref{eq:soundwaves_peak_frequency}, which states that the peak frequency depends linearly on $\beta/H_{\rm PT}$. An interesting prediction of the VLL model under study is the multi--peaked GW signature~\cite{Vieu:2018zze,Ramsey-Musolf:2017tgh,Croon:2018new}, which is generated by the two SFOPTs featured for the points selected by our scans. We find that the GW signature coming from the earlier SFOPT is weaker and peaks at higher frequencies than the one resulting from the later SFOPT. These features are a consequence of having $\alpha_2 < \alpha_1$ and $\left( \beta/H_{\rm PT} \right)_2 > \left( \beta/H_{\rm PT} \right)_1$. 

%%%%%%%%%%%%%%%%%%
\begin{figure}[t!]
  \centering
  \includegraphics[width=0.7\textwidth]{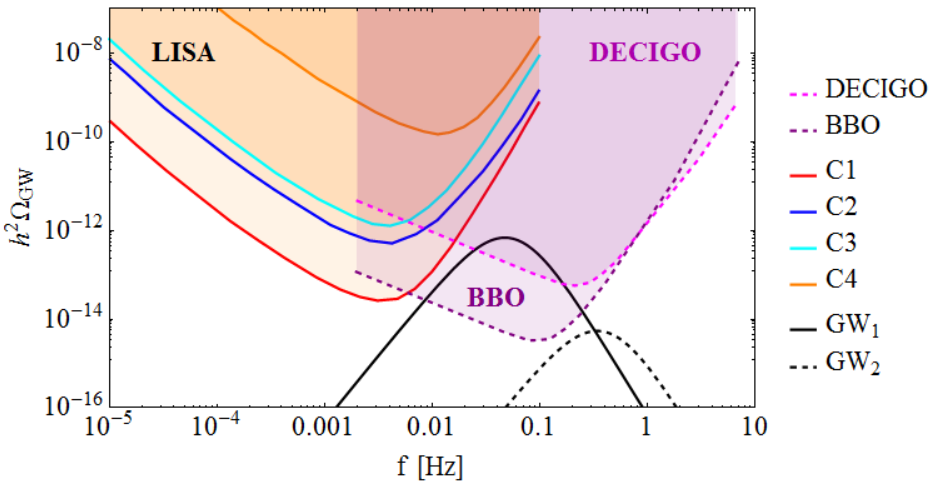}
  \caption{\small GW spectrum $h^2 \Omega_{\rm GW}$ for the earlier (dashed black line) and later (solid black line) SFOPTs, as a function of the frequency $f$ in Hertz, for the benchmark point BM1 described in Table~\ref{tab:table-bm}. The coloured regions correspond to the sensitivities of several future GW expmeriments: the four possible LISA configurations, C1--C4,\cite{Caprini:2015zlo}, and the DECIGO~\cite{Musha:2017usi} and BBO~\cite{Crowder:2005nr} experiments. The sensitivity curves for the latter two have been taken from Ref.~\cite{Moore:2014lga}. }
  \label{fig:gw-spectrum-1}
\end{figure}
%%%%%%%%%%%%%%%%%%

%%%%%%%%%%%%%%%%%%
\begin{figure}[t!]
  \centering
  \includegraphics[width=0.7\textwidth]{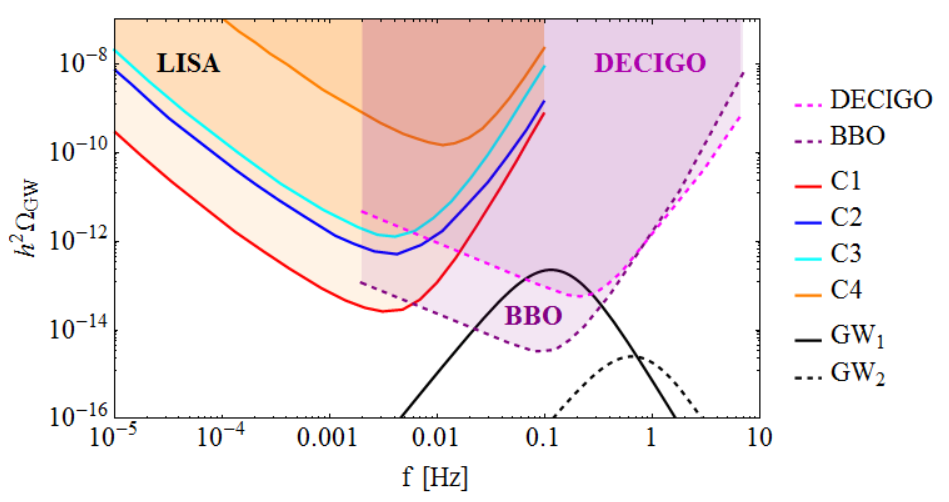}
  \caption{\small Same as Fig.~\ref{fig:gw-spectrum-1}, but for the benchmark point BM2 described in Table~\ref{tab:table-bm}.   }
  \label{fig:gw-spectrum-2}
\end{figure}
%%%%%%%%%%%%%%%%%%

\begin{figure}[t!]
  \centering
  \includegraphics[width=0.7\textwidth]{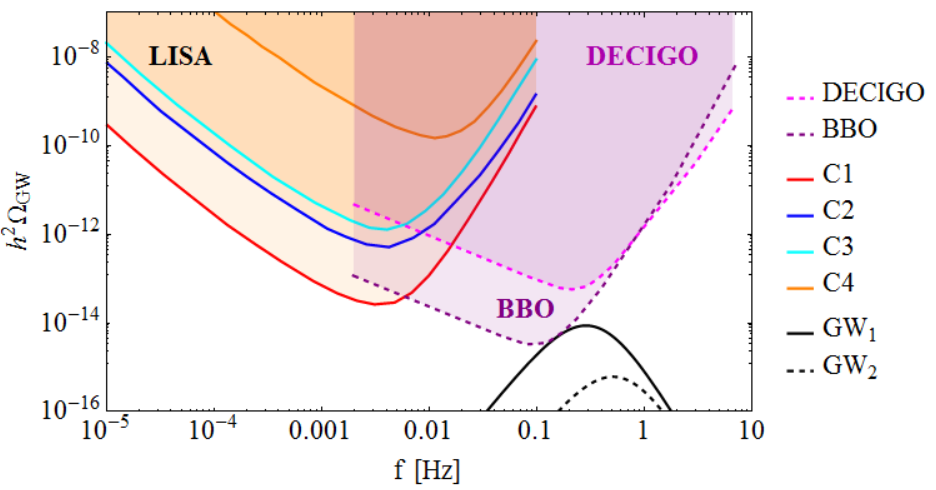}
  \caption{\small Same as Fig.~\ref{fig:gw-spectrum-1}, but for the benchmark point BM3 described in Table~\ref{tab:table-bm}.   }
  \label{fig:gw-spectrum-3}
\end{figure}

The parameters relevant to the GW spectrum are listed in Table.~\ref{tab:table-gw}. We show in Fig.~\ref{fig:gw-spectrum-1} - Fig.~\ref{fig:gw-spectrum-3} the spectrum $h^2 \Omega_{\rm GW}$, as a function of the frequency $f$, of the GWs released during the two SFOPTs predicted for the three benchmarks listed in Table~\ref{tab:table-bm}. The solid black line corresponds to the later PT, while the dashed one stands for the earlier PT. Owing to the fact that $\beta_1 < \beta_2$ and $\alpha_1 > \alpha_2$, the GW signal from the later (and stronger) PT is generally 2--3 orders of magnitude stronger than the signal from the earlier PT. The GW$_{1}$ signal, corresponding to the later PT, peaks at frequencies of $\sim 0.05$~Hz, $\sim 0.1$~Hz, and $\sim 0.3$~Hz for the three benchmark points respectively, while the other signal's peak is located at $\sim 0.4$~Hz, $\sim 0.7$ ~Hz, and $\sim 0.5$ ~Hz for the three benchmark points respectively. The colored regions represent the future sensitivity of the LISA experiment for four possible configurations, C1--C4~\cite{Caprini:2015zlo}, and of the DECIGO~\cite{Musha:2017usi} and BBO~\cite{Crowder:2005nr} experiments. The sensitivity curves are taken from Ref.~\cite{Caprini:2015zlo} for LISA and from Ref.~~\cite{Moore:2014lga} for DECIGO and BBO. We observe that, even if the GW$_1$ signal has a strength comparable to the sensitivity of the C1 configuration of LISA, it peaks at a frequency which does not correspond to the maximum LISA sensitivity, which lies in the mHz range. Therefore, we conclude that this GW signal would not be detectable by the LISA experiment. However, the stronger GW$_1$ signals, such as the ones from benchmarks BM1 and BM2, can be detected by the DECIGO and BBO experiments, whose projected sensitivities are maximized close to the peak frequency of GW$_1$. On the other hand, the weaker GW$_2$ signal would escape detection at all three GW experiments we consider in this work.

\begin{table}[t]
\renewcommand{\arraystretch}{1.6}
\centering
\begin{tabular}{|c|c|c|c|}
\hline 
&BM1 & BM2 & BM3 \\ \hline
$T_{PT_2}$ (GeV) & 165 &186 & 164\\ \hline
$\alpha_2$ & 0.012 &0.010 & 0.005\\ \hline
$\left( \frac{\beta}{H_{\rm PT}} \right)_{\!2} $ & 6480 &10880 &9690 \\ \hline
 $T_{PT_1}$ (GeV) & 82.8 &97.7 &118 \\ \hline
$\alpha_1$ & 0.074 & 0.060& 0.016\\ \hline
$\left( \frac{\beta}{H_{\rm PT}} \right)_{\!1} $ & 1834 & 3740&7710 \\ \hline
\end{tabular}
\caption{\small Relevant parameters to the GW spectrum.}
\label{tab:table-gw}
\end{table}	 

%%%%%%%%%%%%%%%%%%
%%%%%%%%%%%%%%%%%%

\subsection{LHC signatures}

The main VLL production mechanism at the LHC is pair production. The total VLL pair production cross sections for our benchmarks are of $\mathcal{O}(0.1)$ - $\mathcal{O}(0.4)$~fb at the LHC, whereas the SM fermion + VLL ones are severely suppressed by the small VLL--$\tau$ mixing (see Appendix). For the first two benchmarks, because of $m_{E_1} > m_{N_1} + m_W$, the $E_1 \to N_1 W$ decay channel is open. As it is not suppressed by SM--VLL mixing, it is the dominant decay mode of $E_1$. This also explains the sizeable difference between the widths of $E_1$ and $N_1$. For the third benchmark, the approximate degeneracy of $E_1$ and $N_1$ drastically reduces the $N_1 \to E_1 W$ branching ratio. As a result, $E_1$ has a width of $\sim 30$~MeV (instead of $\mathcal{O}(10)$~GeV in the previous two examples), its most probable decay channel being $E_1 \to \tau h$, with subleading $\nu W$ and $\tau Z$ branching ratios.

\begin{table}[t]
\renewcommand{\arraystretch}{1.6}
\centering
\begin{tabular}{|c|c|c|c|}
\hline 
& BM1 & BM2 & BM3 \\ \hline
$m_{N_1}$~(GeV)	& 400  &401 &466 \\ \hline
$\Gamma_{N_1}$ ~(MeV) & 1.32 &0.38 & 0.96\\ \hline
$m_{E_1}$~(GeV)	& 592  &740 &460 \\ \hline
$\Gamma_{E_2}$ ~(GeV) & 7.46 & 11&0.032 \\ \hline
${\rm BR}(E_1 \to N_1 W)$ & 0.995 & 0.996& - \\ \hline
${\rm BR}(E_1 \to \tau h)$	& $3.5 \times 10^{-3}$  &$2.5 \times 10^{-3}$ &0.606 \\ \hline
  ${\rm BR}(E_1 \to \nu W)$	& $7.6 \times 10^{-4}$  & $9.6 \times 10^{-4}$& 0.29\\ \hline
  ${\rm BR}(E_1 \to \tau Z)$	& $4.6 \times 10^{-4}$ & $4.1 \times 10^{-4}$& 0.104\\ \hline
  $\sigma(pp \to E_1 E_1)$ (fb) & 0.32 &0.13 &0.41 \\ \hline
   $\sigma(pp \to E_1 N_1) (fb)$ & 0.36 &0.09 &0.09 \\ \hline
   $\sigma(pp \to N_1 N_1) (fb)$ & 0.31&0.11 & 0.08\\ \hline
  $\sigma(pp \to f_{\rm NP}f_{\rm SM})$ (fb) & $\mathcal{O}(10^{-4})$ &   $ < 10^{-4}$&$ < 10^{-3}$\\ \hline
\end{tabular}
\caption{\small Predictions for the masses, decay widths, branching ratios, and 13 TeV LHC production cross sections of the lighter VLL eigenstates for the three benchmark points. Here, $f_{\rm NP}$ denotes the new fermions $N_1$ or $E_1$, while $f_{\rm SM}$ stands for a $\tau$ or $\nu$.}
\label{tab:table-collider}
\end{table}	 
 
These findings, alongside with other collider predictions, are collected in Table~\ref{tab:table-collider} for all three benchmarks. In summary, the cross section for VLL pair production at a 13~TeV proton--proton collider are below the fb level, rendering direct searches at the LHC challenging. Moreover, single VLL production (in association with a SM lepton) is below the ab level, which means that such a process is undetectable at the LHC. However, as explained at the end of Sec.~\ref{subsec:gen_predictions}, a more promising search avenue is measuring the Higgs to diphoton signal strength, $\mu_{\gamma\gamma}$, for which our model predicts a rather significant enhancement (but still within experimental limits) with respect to its SM value.

\section{Summary and conclusions}
\label{sec:conclusions}

In this work, we have studied the impact of a minimal Dirac VLL model on the thermal history of the Universe. We have shown that, indeed, TeV-scale VLLs can induce strongly first order EW phase transitions, which would generate favorable conditions for a dynamical origin of the baryon asymmetry in the Universe. We have also discussed the collider and GW predictions of such a scenario, and assessed how it can be tested at the LHC and at future GW experiments, such as LISA. 

Remarkably, such a simple setup predicts a complex phase structure of the Universe, involving three PTs: a crossover in the very early Universe ($T \gtrsim 500$~GeV), which results in EWSB, followed by two SFOPTs, both at EW--scale temperatures, the first one restoring EW symmetry, and the last one breaking it again. This non-trivial succession of PTs can be traced back to competing thermal and non--thermal effects coming from the VLLs and the SM contribution. To the best of our knowledge, our model provides a first example of single--field multistep strongly first order EW phase transitions.

Since our model exhibits two SFOPTs, it also predicts a GW signature featuring two peaks. We noticed that, for nearly all the points from our scan, the later SFOPT produces a stronger GW signal and peaks at a lower frequency than the earlier one. Generally, the signal peaks are located at frequencies in the $0.01-1$~Hz range. Meanwhile, the maximum LISA sensitivity (corresponding to the C1 scenario) is achieved for $f\sim 4$~mHz, whereas DECIGO and BBO are most sensitive to $f \sim 0.1-0.3$~Hz. Thus, even if for some points of parameter space our model feature a later PT with a GW signal strength comparable to the reach of LISA, the offset between the GW peak frequency and the frequency of maximum LISA sensitivity leads us to conclude that LISA is unlikely to detect the GW signature predicted by our model. However, the detection prospects at DECIGO and BBO are more optimistic, as their typical maximum sensitivity is achieved at frequencies close to the peak frequencies of the GW$_1$ signal.

Finally, on the collider front, direct production of the VLLs is not a promising way of testing our model at the LHC, as the predicted cross--sections of VLL pair--production are around the $0.1$~fb level. Even without a dedicated collider analysis, we estimate that such low rates would make pair--produced VLLs difficult to study at the LHC. Furthermore, production of a VLL in association with a SM lepton has a cross section of $\mathcal{O}(10^{-4})$~fb, which is beyond doubt out of the reach of LHC. Instead, the measurement of the diphoton Higgs signal strength, $\mu_{\gamma\gamma}$, is a powerful collider probe of our scenario. The relatively light charged VLLs which, as required for inducing a SFOPT in the Early Universe, couple strongly to the Higgs, enhance $\mu_{\gamma\gamma}$ by at least $15\%$ with respect to its SM value of 1. While the current $\mu_{\gamma\gamma}$ searches still allow for such departures from the SM, the high--luminosity (HL) option of LHC will constrain $\mu_{\gamma\gamma}$ at the level of $5\%$~\cite{ATLAS:2013hta,CMS:2013xfa,ATL-PHYS-PUB-2014-016}. Therefore, we conclude that, through $\mu_{\gamma\gamma}$, HL--LHC will be able to fully test our scenario of VLL-induced SFOPTs.

\section*{Acknowledgments}
\label{sec:acknowledgments}
We thank K.S.~Babu, J.~Braathen, D.~Egana-Ugrinovic, I.~Lewis, A.~Morais, K.~Sinha, D.~Teresi, and C.~Wagner for helpful discussions. This work is supported by University of Nebraska-Lincoln, National Science Foundation under grant number PHY-1820891, and the NSF Nebraska EPSCoR under grant number OIA-1557417.

\appendix

\section{Calculation of Gravitational Wave Spectrum} \label{appendix-gw}
\renewcommand{\theequation}{\thesection.\arabic{equation}}
\setcounter{equation}{0}

Our calculation of the GW spectrum relies on the results collected in Ref.~\cite{Caprini:2015zlo}, while the notation follows the one from Ref~\cite{Huang:2016cjm}.

We start by calculating the $\alpha$ and $\beta$ parameters, which have been already discussed in the main text. In order to determine their values, several computational steps need to be taken. First, the ``bounce'' action, $S_3(T)$, has to be evaluated (see e.g. Ref.~\cite{Dine:1992wr}):
\begin{equation}
S_3(T) = 4 \pi \int d r \, r^2 \left[\frac{1}{2} \left( \frac{d \phi_b}{d r}\right)^2 + V(\phi_b(r),T) \right],
\label{eq:bounce_action}
\end{equation} 
with $\phi_b(r)$ being the $SO(3)$--symmetric bounce solution, which describes the field value profile of an expanding spherically symmetric bubble, with $r$ measuring the distance from the center of the bubble. For a given temperature $T$, the bounce solution $\phi_b (r)$ satisfies the following differential equation:
\begin{equation}
\frac{d^2 \phi_b}{d r^2} + \frac{2}{r}\frac{d \phi_b}{d_r} = V^{\prime} (\phi,T), \quad {\rm with} \quad \phi_b (r=\infty) = \phi_{\rm true} \quad {\rm and} \quad \frac{d \phi_b}{dr}\bigg|_{r=0} = 0,
\label{eq:bounce_solution_diff_eq}
\end{equation}
where $\phi_{\rm true} $ represents the field coordinate of the true vacuum (the global minimum of the potential at a given temperature) and the prime symbol denotes differentiation with respect to $\phi$. 

At finite temperature, the nucleation rate of true vacuum bubbles behaves as $ \Gamma_n \sim A(T)  \exp (-\frac{S_3(T)}{T})$ (see e.g. Ref.~\cite{Marzo:2018nov}). Denoting by $H$ the Hubble expansion rate, the phase transition begins when $\Gamma_n \simeq H$, this condition being equivalent to $\frac{S_3 (T_n)}{T_n} \simeq 142$, which serves as a definition of the nucleation temperature, $T_n$. The PT continues until a fraction of order unity of the universe is filled by true vacuum bubbles, which occurs at a temperature denoted by $T_p$ (percolation temperature). Then, as the bubbles collide, the energy stored in their walls is transferred to the plasma as heat, causing the plasma to reheat to a temperature $T_{\rm reh} > T_p$.  Following Ref.~\cite{Huang:2016cjm}, we make the simplifying approximation that $T_n \simeq T_p \simeq T_{\rm reh} \equiv T_{\rm PT}$, which is justified \textit{a posteriori} by the high values of $\beta/H_{\rm PT}$ that imply a short duration of the PT. We thus define the temperature at which the PT occurs as
\begin{equation}
\frac{S_3 (T_{\rm PT})}{T_{\rm PT}} = 142.
\label{eq:T_PT_definition}
\end{equation}  

The temperature at which the PT occurs, $T_{\rm PT}$, is an essential ingredient for computing the $\alpha$ and $\beta$ parameters, whose physical meaning was described at the beginning of this appendix. Mathematically, the two parameters are defined as~\cite{Ellis:2018mja}
\begin{gather}
\alpha = \frac{\left[ V(\phi_{\rm false},T_{\rm PT}) - V(\phi_{\rm true},T_{\rm PT}) \right] + T_{\rm PT} \left[ \partial_T V(\phi_{\rm true},T) -\partial_T V(\phi_{\rm false},T) \right]_{T_{\rm PT}}}{\rho_{\rm rad, true} (T_{\rm PT})} , \label{eq:alpha_definition} \\ 
\frac{\beta}{H_{\rm PT}} = T_{\rm PT} \frac{d}{d T} \left( \frac{S_3 (T)}{T} \right) \bigg|_{T_{\rm PT}},
\label{eq:beta_definition}
\end{gather}
with $H_{\rm PT}$ being the Hubble rate corresponding to $T_{\rm PT}$, and $\rho_{\rm rad, true} (T_{\rm PT})$ the radiation energy density in the true vacuum phase, $\rho_{\rm rad, t} (T_{\rm PT})= (\pi^2/30) g_{\rm eff,PT} T_{\rm PT}^4$, where $g_{\rm eff,PT}$ denotes the number of relativistic degrees of freedom (again in the true vacuum phase) at temperature $T_{\rm PT}$. $\phi_{\rm false}$ is the $\phi$ coordinate of the false vacuum, while $\phi_{\rm true}$ has the same meaning as in Eq.~\eqref{eq:bounce_solution_diff_eq}. For the SFOPTs present in our model, either $\phi_{\rm false}$ or $\phi_{\rm true}$ are equal to 0.

Having defined $\alpha$ and $\beta$, we now concentrate on the calculation of the GW spectrum. As discussed in Sec.~\ref{subsec:gw-signature}, there are three sources of GWs produced at a SFOPT: bubble collisions, sound waves in the plasma, and magnetohydrodynamic (MHD) turbulence. As in Ref.~\cite{Caprini:2015zlo}, we suppose that the three contributions to the stochastic GW background combine linearly, giving the total GW signal:
\begin{equation}
h^2 \Omega_{\rm GW} \simeq h^2 \Omega_{\rm col} + h^2 \Omega_{\rm sw} + h^2 \Omega_{\rm turb}.
\label{eq:GW_source_linear_superposition}
\end{equation}

The spectrum of the GWs produced by bubble collisions reads
\begin{equation}
h^2 \Omega_{\rm col} (f) = 1.67 \times 10^{-5} \left( \frac{0.11 v_w^3}{0.42 + v_w^2}\right) \left(\frac{\beta}{H_{\rm PT}}\right)^{-2} \left(\frac{\kappa_{\rm col} \alpha}{1+\alpha}\right)^2 \left( \frac{g_{\rm eff, PT}}{100} \right)^{-1/3} \frac{3.8 \left( f/f_{\rm col} \right)^{2.8}}{1+2.8\left( f/f_{\rm col} \right)^{3.8}},
\label{eq:collision_gw_spectrum}
\end{equation}
where $\kappa_{\rm col}$ is the fraction of latent heat converted into bubble wall kinetic energy, and $v_w$ the bubble wall speed. The bubble collision spectrum has a peak frequency given by
\begin{equation}
f_{\rm col} = \left( 1.65\times 10^{-5} \, {\rm Hz} \right) \left( \frac{0.62}{1.8 - 0.1 v_w + v _w^2} \right) \left( \frac{\beta}{H_{\rm PT}} \right) \left( \frac{T_{\rm PT}}{100 \, {\rm GeV}} \right) \left( \frac{g_{\rm eff,PT}}{100} \right)^{1/6}.
\label{eq:collision_peak_frequency}
\end{equation}  
The sound wave contribution is given by
\begin{equation}
h^2 \Omega_{\rm sw}(f) = 2.62 \times 10^{-6} \, v_w \left(\frac{\beta}{H_{\rm PT}}\right)^{-1} \left(\frac{\kappa_v \alpha}{1+\alpha}\right)^2 \left( \frac{g_{\rm eff,PT}}{100} \right)^{-1/3} \frac{7^{3.5}\left( f/f_{\rm sw} \right)^3}{\left( 4 + 3 \left( f/f_{\rm sw} \right)^2 \right)^{3.5}}
\label{eq:soundwaves_gw_spectrum},
\end{equation}
and peaks at a frequency
\begin{equation}
f_{\rm sw} = \left( 1.9 \times 10^{-5} \, {\rm Hz} \right) \left( \frac{1}{v_w} \right) \left( \frac{\beta}{H_{\rm PT}} \right) \left( \frac{T_{\rm PT}}{100 \, {\rm GeV}} \right) \left( \frac{g_{\rm eff,PT}}{100} \right)^{1/6}.
\label{eq:soundwaves_peak_frequency}
\end{equation}
In Eq.~\eqref{eq:soundwaves_gw_spectrum}, $\kappa_v$ stands for the fraction of latent heat transferred into the bulk motion of the plasma. Finally, the MHD turbulence decay contributes to the GW spectrum as 
\begin{equation}
h^2 \Omega_{\rm turb} (f) = 3.35 \times 10^{-4}  v_w  \left(\frac{\beta}{H_{\rm PT}}\right)^{-1} \left(\frac{\kappa_{\rm turb} \alpha}{1+\alpha}\right)^{3/2} \left( \frac{g_{\rm eff}}{100} \right)^{-1/3} \!\! \frac{\left( f/f_{\rm turb} \right)^{3}}{\left( 1+f/f_{\rm turb} \right)^{11/3}\left( 1 + 8\pi f/h_{*} \right)},
\label{eq:mhd_gw_spectrum}
\end{equation}
where $\kappa_{\rm turb}$ is the fraction of latent heat transferred to turbulent plasma motion, and 
\begin{align}
f_{\rm turb} &= \left( 2.7\times 10^{-5} \, {\rm Hz} \right) \left( \frac{1}{v_w} \right) \left( \frac{\beta}{H_{\rm PT}} \right) \left( \frac{T_{\rm PT}}{100 \, {\rm GeV}} \right) \left( \frac{g_{\rm eff}}{100} \right)^{1/6}, \notag \\
h_{*} &= \left( 1.65\times 10^{-5} \, {\rm Hz} \right) \left( \frac{T_{\rm PT}}{100 \, {\rm GeV}} \right) \left( \frac{g_{\rm eff}}{100} \right)^{1/6}.
\end{align}

The question of choosing the efficiency factors $\kappa_{\rm col, sw, turb}$ and the bubble wall velocity $v_w$ is model-dependent and involves certain calculations and assumptions regarding the dynamics of the bubble walls. Therefore, such a task is beyond the scope of the current work. Instead, we resort to a much simpler approach. Using the results from Ref.~\cite{Espinosa:2010hh}, in which the authors numerically express $\kappa_v$ as a function of $v_w$~\footnote{In Ref.~\cite{Espinosa:2010hh}, $\kappa_v$ is denoted simply as $\kappa$ and $v_w$ as $\xi_w$.} for different values of $\alpha$, we choose the bubble wall velocity $v_w$ such that it corresponds to the maximum value of $\kappa_v$ for a given value of $\alpha$. In our model, the strongest PTs typically have $\alpha \lesssim 0.1$, and we choose $v_w = 0.6$, from which it follows that $\kappa_v \simeq 0.4$~\cite{Espinosa:2010hh}. Concerning the turbulence efficiency factor, it is given by $\kappa_{\rm turb} = \epsilon \kappa_v$, with the choice $\epsilon = 0.05$~\cite{Huang:2016cjm}. Finally, we take for definiteness $\kappa_{\rm col} = 0.5$, but nevertheless mention that the choice for $\kappa_{\rm col}$ plays little role in our analysis, as the sound wave contribution is the one dominating by far the GW spectrum predicted by our scenario.

\section{SM--VLL Mixing for Collider Phenomenology}
\label{appendix-collider}
\setcounter{equation}{0}

In absence of mixing with the SM fermions, the lightest VLL of our model would be stable. For the range of $N_1$ and $E_1$ masses predicted by our scenario, this would not be a viable option. On the one hand, if $m_{E_1} > m_{N_1}$, then a stable $N_1$ would not be a suitable Dark Matter (DM) candidate: the large Yukawas necessary for a SFOPT would induce a strong $SU(2)_L$--doublet component in $N_1$. This would imply a sizeable $Z N_1 N_1$ coupling, which would be in conflict with null results from DM direct detection experiments~\cite{Angelescu:2016mhl}. On the other hand, for $m_{N_1} > m_{E_1}$, $E_1$ would be a stable charged particle, but we choose to not pursue this possibility. Therefore, in order to avoid stable VLLs, our model has to feature a mixing between the VLL sector and the SM fermions. 

We thus choose to introduce a small $\tau$ lepton--VLL mixing in our model, which is achieved by adding the following Yukawa terms to the Lagrangian in Eq.~\eqref{eq:vll_yuk_lagrangian}:
\begin{equation}
-\mathcal{L}_{\rm mix} = y_1 \, \overline{L}_L H \tau_R + y_2 \, \overline{L}^3_L H E^{\prime}_R + {\rm h.c.} \, ,
\label{eq:sm-vll_mixing_lagrangian}
\end{equation}
where $L^3_L$ is the third generation SM lepton doublet. For simplicity, we suppose that the SM neutrinos do not mix with the neutral VLLs. When presenting our collider predictions for the benchmarks in Sec.~\ref{sec:benchmarks}, we choose $y_1=y_2=0.05$. We have explicitly checked that these values for $y_{1,2}$ predict deviations from the SM values in the $W\tau\nu$ and $Z\tau\tau$ couplings which are below the sensitivity achieved at the LEP experiment~\cite{ALEPH:2005ab} or in $\tau$ lifetime measurements~\cite{Tanabashi:2018oca}. More precisely, the precision in the measurement of the $Z\tau\tau$ axial coupling and the $W\tau\nu$ couplings (from $\tau$ lifetime~\footnote{Since all the (tree level) $\tau$ decays are proportional to the $W\tau\nu$ coupling squared, a rescaling of the latter would change the $\tau$ lepton lifetime accordingly.} measurements) is at the permille level, while the precision for the $Z\tau\tau$ vectorial coupling measurement is at the percent level (see e.g. PDG~\cite{Tanabashi:2018oca}). Meanwhile, in our model, we checked that $\tau$--VLL mixing amounts to shifts with respect to the SM which are typically one order of magnitude below the corresponding experimental sensitivities.

We now briefly discuss the deviation in the $h\tau\tau$ Yukawa coupling induced by $\tau$--VLL mixing. We start by noting that, in order to address this problem, we would in principle need to add another term in the Lagrangian, namely the SM--like term $y_\tau \overline{L}_L^3 H \tau_R + {\rm h.c.}$, with $y_\tau$ a free parameter. However, once the values of all the other Yukawas are specified, $y_\tau$ can be chosen such that the $\tau$ mass central experimental value, $m_\tau \simeq 1.777$~GeV, is reproduced. Once this step is performed, the result is a deviation in the $h\tau\tau$ physical Yukawa coupling. We have checked that for our values of $y_{1,2}$ and the scanned values of the other parameters in Eq.~\eqref{eq:vll_yuk_lagrangian}, the $h\tau\tau$ coupling never deviates by more than $\sim$5\% from its SM value, which is below the current sensitivity achieved by meaurements of the $h\to\tau\tau$ decay at the LHC~\cite{Aad:2015vsa,Sirunyan:2018koj}.

\section{Theoretical Constraints}
\label{appendix-th-constraints}
\setcounter{equation}{0}

In this Appendix, we briefly overview the theoretical constraints affecting our model. Any model containing new fermions that couple strongly to the Higgs suffers from theoretical inconsistencies, and can only be viewed as a low--energy effective description of a more fundamental UV completion. Indeed, the effect of large Yukawas is two--fold: on the one hand, they tend to push the one--loop effective potential towards negative values for large values of $\phi$; on the other hand, the Renormalization Group Equations (RGEs) drive the strong Yukawas towards Landau poles, which signal the breakdown at high energies of the theory under consideration. Moreover, too high values of the Yukawa couplings can lead to unitarity violation in various processes, such as VLL--VLL scattering. Therefore, in the following, we discuss the problems of vacuum stability, Yukawa Landau poles, and unitarity in our model. For concreteness, we analyze the benchmark point ``BM1'', discussed in Sec.~\ref{sec:benchmarks}. Since BM1 exhibits the strongest later PT among the points in our scan, we expect that the above theoretical consideration will be most constraining for this point, as the other points typically present weaker Yukawa couplings.

We start by considering the issue of vacuum stability. Due to the large Yukawa couplings we consider, loop effects associated with the new fermions destabilize the effective potential at high field values, rendering it unbounded from below. This has catastrophic consequences for the EW vacuum, which becomes unstable. For instance, in the case of benchmark point BM1, the zero--temperature one--loop effective potential becomes negative at $\phi \simeq 2.5$~TeV, which is slightly above the mass scale of the heavier VLL eigenstates, $m_{N_2}\simeq 1.6$~TeV and $m_{E_2}\simeq 1.8$~TeV. 

The simplest solution for stabilizing the EW vacuum is to add the dimension--6 effective operator $\left(H^{\dagger} H \right)^3 / \Lambda^2$~\cite{Egana-Ugrinovic:2017jib}, which can arise, for example, from a UV completion featuring compositeness/strong dynamics. Indeed, we find that, for $\Lambda \leq 3.2$~TeV, the potential increases monotonically for $\phi > v$ and thus is no longer unbounded from below. Therefore, the EW minimum becomes absolutely stable. For the slightly higher value of $\Lambda=5$~TeV, the zero--temperature potential develops a second minimum, deeper than the EW one, at $\phi \sim 6.4$~TeV. In this case, tunneling from the EW minimum to the deeper one becomes possible. We find that the $O(4)$--symmetric bounce action for this transition is $S_4 \simeq 20$, and, using the formalism from Refs.~\cite{Coleman:1977py,Callan:1977pt}, we estimate the EW vacuum lifetime to be roughly 40 orders of magnitude lower than the age of the Universe. Clearly, such a situation is ruled out. For higher values of the cutoff scale, the EW vacuum becomes even more short--lived.

Before moving on, we mention that, as argued in Ref.~\cite{Egana-Ugrinovic:2017jib}, we expect the $\left(H^{\dagger} H \right)^3$ operator to have a negligible contribution (of the order $v^2/\Lambda^2$) on the dynamics of the PTs in our model.

We now turn our attention to the Landau poles appearing in our model. For our RGE analysis, we have adapted the general 1--loop Yukawa RGEs from Ref.~\cite{Machacek:1983fi} to our case, taking into account only the running of the VLL and top Yukawas and neglecting the subleading contribution of the gauge couplings. The resulting beta functions are given by
\begin{align}
32 \pi^2 \beta_{y_t} &= y_t \left( \frac{9}{2} \left|y_t\right|^2 + \left|y_{N_L}\right|^2 + \left|y_{N_R}\right|^2 + \left|y_{E_L}\right|^2 + \left|y_{E_R}\right|^2 \right), \notag \\
32 \pi^2 \beta_{y_{N_L}} &= y_{N_L} \left( \frac{5}{2} \left|y_{N_L}\right|^2 - \frac{1}{2} \left|y_{E_L} \right|^2 + \left|y_{N_R}\right|^2 + \left|y_{E_R}\right|^2 + 3 \left|y_t\right|^2  \right),
\label{eq:yuk-rge}
\end{align} 
with $\beta_y \equiv \frac{d y}{d \log \mu}$. The beta functions for $y_{N_R}$ and $y_{E_{L,R}}$ can be obtained from the second line of Eq.~\eqref{eq:yuk-rge} by an appropriate interchange of the indexes, i.e. $N \leftrightarrow E$ and/or $L \leftrightarrow R$. As an example, interchanging $N$ and $E$ in the expression of $\beta_{y_{N_L}}$ gives $\beta_{y_{E_L}}$. For the initial conditions of the RGEs, we consider a starting energy scale $\mu_0$ and take $y_t (\mu_0) = 1$, whereas for the VLL Yukawas at $\mu_0$ we take the values listed in Table~\ref{tab:table-bm} in the BM1 column. Under these assumptions, we find that the Yukawa Landau poles occur at a scale around $\sim 20 \, \mu_0$. Thus, taking $\mu_0 = 500$~GeV, which corresponds to a scale associated with the lighter VLL mass eigenstates, would lead to a Landau pole at $\mu \sim 10$~TeV. This finding is consistent with a UV--completion at $\Lambda \leq 5$~TeV that would stabilize the potential, as discussed in the previous paragraph.

Finally, we study the partial wave unitarity constraints on the possible values of Yukawa couplings. We focus on the $2 \to 2$ process of  VLL--VLL scattering, as its amplitude scales as the square of VLL Yukawa couplings. We neglect the subleading contributions coming from gauge boson exchanges, and, using the Feynman gauge, we consider only Feynman diagrams involving internal scalars (Higgs and Goldstones). We work in the high--energy limit where all the masses involved in the process can be neglected. More specifically, we consider the full transition matrix of $\psi_i \bar\psi_j \to \psi_k \bar\psi_l $ scattering amplitudes, with $i,j,k,l$ labeling the four possible VLL states, and restrain ourselves to the $++++$ helicity states (see e.g. Ref.~\cite{DiLuzio:2016sur} for a discussion). The $J=0$ (vanishing total angular momentum) partial wave unitarity criterion, applied to the highest eigenvalue of the $\psi_i \bar\psi_j \to \psi_k \bar\psi_l $ transition matrix, leads to the following constraint:
\begin{equation}
{\rm max} \, \left\lbrace  |y_{N_L}|^2 + |y_{E_R}|^2 \, , \, |y_{N_R}|^2 + |y_{E_L}|^2  \right\rbrace < 8\pi,
\label{eq:unitarity_constraint}
\end{equation}
which is automatically satisfied by imposing $|y_{\rm VLL}| < \sqrt{4\pi}$, as we did in our scan.

\newpage

\bibliographystyle{utphys}
\bibliography{ewptref}

\end{document}